\documentclass[10pt,twocolumn]{revtex4-1}
\usepackage{bm}
\usepackage{hyperref}
\usepackage{makeidx}
\usepackage{amsmath}
\usepackage{graphicx}
\usepackage{dcolumn}
\usepackage{comment}
\usepackage{color}
\usepackage{graphics}
\usepackage{amsfonts}
\usepackage{amssymb}
\usepackage{epsfig}
\usepackage{float}
\usepackage{cleveref}
\usepackage[british,UKenglish,USenglish,english,american]{babel}
\usepackage{epstopdf}
\usepackage{graphics, subfigure}
\usepackage{array}
\usepackage{physics}

\begin{document}

\title{Saturation Field as a Direct Probe of Exchange and Single-Ion Anisotropies in Spin-1 Magnets}


\author{M.~A.~R.~\surname{Griffith}$^{1}$}
\email{griffithphys@gmail.com}

\author{S.~\surname{Rufo}$^{2,3}$}
\author{H.~\surname{Caldas}$^{4}$}
\author{F.~Dinola~\surname{Neto}$^{1}$}
\author{Minos~A.~\surname{Neto}$^{1}$}
\author{J.~R.~\surname{Viana}$^{1}$}

\affiliation{$^{1}$Departamento de Física, Universidade Federal do Amazonas, 
Av.\ Gen.\ Rodrigo Octávio 6200, Coroado~I, 69080-900 Manaus, Brazil}

\affiliation{$^{2}$Instituto de computação, Universidade Federal do Amazonas, 
Av.\ Gen.\ Rodrigo Octávio 6200, Coroado~I, 69080-900 Manaus, Brazil}

\affiliation{$^{3}$Instituto Militar de Engenharia IME, 
Praça Gen. Tibúrcio, 80 - Urca, Rio de Janeiro - RJ, 22290-270, Brazil}

\affiliation{$^{4}$Departamento de Ciências Naturais, Universidade Federal de São João del Rei,
Praça Dom Helvécio 74, 36301-160 Sao João del Rei, MG, Brazil}

\date{\today }

\begin{abstract}
High magnetic fields provide a direct route to probe the anisotropies that govern spin dynamics in layered magnets. 
Using the SU(3) bond–operator framework for spin–1 systems, we derive analytic expressions for the magnon spectrum and the critical fields delimiting the field–induced ordered phase. 
We show that the upper critical field $h_{c2}$ carries a simple and quantitative fingerprint of both exchange anisotropy and single–ion symmetry breaking, enabling high–field experiments to serve as sensitive probes of microscopic anisotropy. 
We further map how these anisotropies, together with interlayer coupling, control the extent and location of the magnon Bose–Einstein condensation dome. 
Our results provide experimentally accessible criteria for identifying symmetry–breaking mechanisms in real spin–1 materials.
\end{abstract}

\maketitle

\section{Introduction}

The discovery of materials exhibiting quasi-two-dimensional magnetic properties has enabled the experimental validation of several theoretical models for these systems. Their study has deep implications for understanding quantum materials, magnetism, and even high-temperature superconductivity\cite{CupratosRevModPhys.66.763}. 

The interplay between interlayer interactions and single-ion anisotropy in spin-1 systems provides a fertile ground for exploring complex quantum phases, unconventional magnetic ground states, and quantum critical phenomena—particularly in quasi-two-dimensional or layered magnetic materials.

Quasi-two-dimensional models are defined by the presence of weak interlayer coupling \cite{Rogado2002}. In general, the interlayer interaction stabilizes the long-range order at finite temperatures, qualitatively changing the quasi-two-dimensional magnetic behavior compared to pure 2D models~\cite{RevModPhys.78.17}.

In addition to interlayer interactions in two-dimensional magnetic materials, single-ion anisotropies play a crucial role in stabilizing long-range antiferromagnetic (AFM) order, as they introduce directional preferences for spin orientation~\cite{Caci2021}. Many layered spin-1 materials exhibit significant single-ion anisotropy due to their crystal-field environment, particularly when transition-metal ions such as Ni$^{2+}$, Fe$^{2+}$, or Mn$^{2+}$ are involved~\cite{Hamer2010, Manson2023}.

Several quasi-two-dimensional spin-1 compounds have been identified in recent years, including $BaNi_2V_2O_8$, $CsNiCl_2$ and $La_2NiO_4$ \cite{Knafo,Klyushina2017,PhysRevB.96.214428}. These materials exhibit both interlayer coupling and single-ion anisotropy. In particular, the $La_2NiO_4$ is a layered perovskite.

Their physical structure arises for alternating layers of $LaO$ and $NiO_2$, which form a natural quasi‑2D geometry. Beyond that, the $NiO_2$ planes are square lattices of $Ni^{2+}$ (spin‑1) ions, responsible for magnetic properties \cite{PhysRevResearch.5.033113}. $LaO$ layers separate $NiO_2$ planes, resulting in weak interlayer magnetic coupling, making it a quasi‑2D antiferromagnet. 

Although the physical properties of purely one-dimensional spin-1 systems are well understood, the anisotropic quasi-two-dimensional cases in the presence of a magnetic field remains less explored. 
There are fundamental questions in the context of multicritical behavior, nematic and multipolar order (absent in spin-½ systems) \cite{PIRES20111977,PAPANICOLAOU}, and the dimensional crossover from 2D to 3D magnetic order in the presence of anisotropies and magnetic field.  Furthermore, under an external magnetic field, magnons may undergo a quasi–Bose-Einstein condensation (BEC), whose nature is influenced by the weak interlayer coupling between planes and the anisotropic character of the system \cite{PhysRevB.79.054431}. The BEC can be interpreted as a spontaneous transverse magnetic order in the $xy$ plane for antiferromagnetic spin-1 systems.

Further interplay coupling, the crystal-field distortions generate
orbital asymmetries that, through spin--orbit coupling, produce both the
dominant axial single--ion term $D(S^z)^2$ and the weaker in-plane component
$D_x(S^x)^2$, while simultaneously inducing inequivalent transverse and
longitudinal superexchange channels ($J_x=J_y=J$ and $J_z\ne J$). These effects jointly reflect the
spin--orbit–driven projection of lattice anisotropy onto the spin sector.
High magnetic fields provide a direct probe of the anisotropies that govern spin dynamics in layered magnets. We show that the saturation field $h_{c2}$ encodes a quantitative fingerprint of both exchange anisotropy and single-ion symmetry breaking, allowing high-field measurements to be used as a sensitive diagnostic of microscopic anisotropy. We also analyze how these anisotropies and interlayer couplings affect the field-induced ordering transition in spin-1 systems, within an SU(3) bond-operator framework that yields a controlled description of the ground state and the associated magnon condensation.

The paper is organized as follows. In Sec.~\ref{Model} we introduce the minimal
Hamiltonian used to characterize the magnetic anisotropies. Section~\ref{Method}
presents the $SU(3)$ Schwinger boson formalism and the resulting variational
equations. The main results and their implications are discussed in
Sec.~\ref{Results}. Finally, Sec.~\ref{Conclusion} summarizes our conclusions
and outlines possible extensions.
\vspace{-0.1cm}
\section{Model} \label{Model}
To capture the competing sources of magnetic anisotropy relevant to layered
spin systems, we consider a minimal spin-1 model that incorporates both
exchange anisotropy and single–ion terms associated with axial and in–plane
distortions. The corresponding Hamiltonian can be as
\begin{align}\label{Model1}
 \mathcal{H}&= J\sum_{\langle i,j \rangle} \bm{\hat{S}}_{i} \bm{\hat{S}}_{j}  + J_{z} \sum_{\langle i,j \rangle} \bm{\hat{S}}_{i} \bm{\hat{S}}_{j}+D\sum_{i} (\hat{S}^{z}_{i})^{2}  \nonumber \\
&+D_{x}\sum_{i} (\hat{S}^{x}_{i})^{2} -g \mu_B B \sum_{i} \hat{S}^{z}_{i},
\end{align}
where, $J$ denotes the nearest-neighbor exchange interaction, while $J_z$ represents the interlayer coupling. $D$ controls the easy-plane single-ion anisotropy and $D_x$ parametrizes the single-ion distortion anisotropy. The magnetic field $B$ (perpendicular to the $xy$-plane) couples to the spins via the spectroscopic Landé factor $g$ and the Bohr magneton $\mu_B$. The spin interaction term has the XXZ form,
$\hat{\mathbf{S}}_{i}\hat{\mathbf{S}}_{j}
= S_i^{x} S_j^{x} + S_i^{y} S_j^{y} + R\, S_i^{z} S_j^{z}$,
 where $R$ parametrizes the exchange anisotropy. The regime $0 \le R < 1$ corresponds to an easy-plane anisotropy, in which the $xy$ components of the spins are energetically favored over their $z$ components. This minimal description retains only the essential ingredients that break
spin-rotational symmetry, allowing us to track how each anisotropy contributes to the high-field behavior. As we show below, the resulting upper critical field $h_{c2}$ provides a clear and quantitative signature of these anisotropy mechanisms.

The Hamiltonian Eq.~(\ref{Model1}) exhibits a rich realm of ground states. The limit $D_x = 0$, $J^z = 0$ and $J_z =1$, was studied by Wang and Wang~\cite{Wang,WanglongPhysRevB}. When $D < D_C$, the system enters into the ordered Néel phase. As shown by Hirsch and Tang \cite{TangPhysRevB.40.4769}, in the Schwinger boson formalism, the condensation of bosons leads to a phase transition to the ordered phase. In the antiferromagnetic ordered phase, the self-consistent harmonic approximation (SCHA) shows that the dispersion relation consists of two nondegenerate branches: one gapless, the other one gapped for $D < D_C$ (the gap vanishes at $D = D_C$).





Here, drastic changes occur in the Hamiltonian Eq.~(\ref{Model1}) as $D$ is tuned from very small to very large values~\cite{Wang}. For $D$-Large than a critical value $D_C$, where a quantum phase transition takes place, the phase consists of a unique ground state with total magnetization $S^z_{total}=0$ separated by a gap from the first excited states, which lies in the sectors $S^z_{total}=\pm 1$.

For positive values of $D \le D_C$, the system resides in a gapless Néel phase at zero temperature.
Zhang \textit{et al}.~\cite{Zhang} pointed out that the interplay between Heisenberg exchange and single ion anisotropies with external magnetic field can result in a variety of quantum phases. In the regime $J_z = 0$ and with an in-plane magnetic field $h$, Eq.~(\ref{Model1}) was previously studied by Pires \textit{et al.}~\cite{Pires1} via the SCHA. As is well known, a finite magnetic field $h$ can give rise to a range of magnetic phenomena both in one~\cite{NPB2009} and two dimensions~\cite{PhysRevB.80.115428}. This standard approach yields an accurate description of the ordered phase, even when an external field is applied in the $xy$ plane. In contrast, it becomes unreliable when the magnetic field is oriented along the $z$ direction. Here, we reexamine the model with the field perpendicular to the $xy$ plane. A field along $z$ reduces the spin gap linearly in $h$, and in this regime, the $SU(3)$ Schwinger boson representation provides a more suitable framework. Although this method is not designed to
capture the Kosterlitz–Thouless transition, it accurately describes the order–disorder transition and, as we show below, naturally incorporates the physics of magnon Bose–Einstein condensation in the magnetically ordered phase.

\section{SU(3) Schwinger boson formalism}\label{Method}

The mapping between bosonic and spin operators in the $SU(3)$ Schwinger boson formalism is (see Appendix~\ref{appSU3})
\begin{align}\label{map}
S^+ &= \sqrt{2}(t_z^\dagger d + u^\dagger t_z), \nonumber \\
S^- &= \sqrt{2}(t_z^\dagger u + d^\dagger t_z), \nonumber \\
S^z &= u^\dagger u - d^\dagger d.
\end{align}

As a starting point, we assume a condensation of the $t$ bosons, i.e., $\langle t_z \rangle = \langle t_z^\dagger \rangle = t$, which provides a good ansatz in the regime $D > D_c$. In this framework, $t$ plays the role of a variational parameter. Substituting Eq.~\eqref{map} into the Hamiltonian \eqref{Model1}, we obtain:
{\allowdisplaybreaks
\begin{align}
H &= \frac{J}{2} \sum_{\vec{r}, \vec{\delta} } \Big[ t^2 (d_{\vec{r}}^\dagger d_{\vec{r}+\vec{\delta}} + u_{\vec{r}}^\dagger u_{\vec{r}+\vec{\delta}} + d_{\vec{r} }u_{\vec{r}+\vec{\delta}}+u_{\vec{r}}d_{\vec{r}+\vec{\delta}}  + \text{H.c.}) \nonumber \\
&+ R(u_{\vec{r}}^\dagger u_{\vec{r}} - d_{\vec{r}}^\dagger d_{\vec{r}})(u_{\vec{r}+\vec{\delta}}^\dagger u_{\vec{r}+\vec{\delta}} - d_{\vec{r}+\vec{\delta}}^\dagger d_{\vec{r}+\vec{\delta}} \,) \Big] \nonumber\\
\displaybreak \nonumber\\
& +  \frac{J^z}{2} \sum_{\vec{r}, \vec{\xi} } \Big[ t^2 (d_{\vec{r}}^\dagger d_{\vec{r}+\vec{\xi}} + u_{\vec{r}}^\dagger u_{\vec{r}+\vec{\xi}} + d_{\vec{r} }u_{\vec{r}+\vec{\xi}}+u_{\vec{r}}d_{\vec{r}+\vec{\xi}}  + \text{H.c.}) \nonumber\\
&+ R(u_{\vec{r}}^\dagger u_{\vec{r}} - d_{\vec{r}}^\dagger d_{\vec{r}})(u_{\vec{r}+\vec{\xi}}^\dagger u_{\vec{r}+\vec{\xi}} - d_{\vec{r}+\vec{\xi}}^\dagger d_{\vec{r}+\vec{\xi}} \,) \Big] \nonumber \\
&+  D \sum_{\vec{r}} \Big [u_{\vec{r}}^\dagger u_{\vec{r}} + d_{\vec{r}}^\dagger d_{\vec{r}} \Big] \nonumber \\
&+ D_{x} \sum_{\vec{r}}  \Big  [\frac{t^2}{2} \Big  [u^\dagger_{k}  u_{k} + d_{k}^{\dagger} d_{k} 
+ u_{k} d_{-k} + u_{k}^\dagger d_{-k}^{\dagger} + H.c. \Big] \nonumber \\
&- h  \sum_r \big [ u_{\vec{r}}^\dagger u_{\vec{r}} - d_{\vec{r}}^\dagger d_{\vec{r}} \Big ] - \sum_{\vec{r}} \mu \big [ u_{\vec{r}}^\dagger u_{\vec{r}}+ d_{\vec{r}}^\dagger d_{\vec{r}} + t^{2} - 1 \Big] \nonumber \\
& + N D_xt^2
\label{hamilt2}
\end{align}
}
%
in which, $\sum_{\vec{r},\vec{\delta}}$ runs over all in-plane nearest-neighbor pairs within each layer, $\vec{\delta}$ spans the set of intralayer nearest-neighbor vectors and $N$ denotes the number of lattice sites. 
By comparison, $\sum_{\vec{r},\vec{\xi}}$ runs over nearest-neighbor interlayer pairs along the $z$ direction, with $\vec{\xi}$ connecting sites in adjacent layers. 
A temperature-dependent chemical potential $\mu_r$ was introduced to impose the local constraint $  u_r^+ u_r + d_r^+ d_r + t_z^+ t_z  = 1$. In practice, $\mu_r$ is replaced by a global chemical potential $\mu$ and the Zeeman term is written as $g \mu_B B = h$.



By performing the mean-field decoupling separately for the intra- and interlayer terms (see Appendix~\ref{appendix:meanfield}) and discarding the anomalous contributions arising from $D_x$ (see Appendix~\ref{Approximations}), we obtain the following diagonal Hamiltonian after a Fourier--Bogoliubov transformation~\cite{fetter_walecka}:
\begin{equation}
H = \sum_k [\omega_k^+ \alpha_k^+ \alpha_k + \omega_k^- \beta_k^+ \beta_k] + C,
\end{equation}
with
\begin{equation}\label{ground}
\omega_k^{(1)} = \omega_k  - h + \frac{1}{2}R (z_1 +  \alpha z_2) m ,
\end{equation}
\begin{equation}
\omega_k^{(2)} = \omega_k  + h - \frac{1}{2}R (z_1 +  \alpha z_2) m ,
\label{1stEx}
\end{equation}
\begin{equation} \omega_k=\sqrt{\Lambda^2_k  - \Delta_k^2}
\end{equation}
\begin{align}
\Lambda_k &= -\mu + D + t^2 D_x  +\frac{R}{2}( z_1+\alpha z_2)(1-t^2) \nonumber \\ 
&+\big(z_1 \gamma^{(1)}_k  + \alpha z_2 \gamma^{(2)}_k\big)t^2 
\end{align}
\begin{align}
\Delta_k &= t^2 D_x +t^2 \big(z_1  \gamma^{(1)}_k  + \alpha z_2  \gamma^{(2)}_k\big) \\  \nonumber 
& -R( z_1p_1\gamma^{(1)}_k+\alpha z_2 p_2\gamma^{(2)}_k) 
\end{align}
%
%
\begin{equation}
\gamma^{(1)}_k = \frac{1}{2} (\cos{(k_x a)}+\cos{(k_y a)}),
\end{equation}
\begin{equation}
\gamma^{(2)}_k =  \cos{(k_z a)} ,
\end{equation}
\begin{align}
C &=  \mu N  (1-t^2) -\frac{1}{4} N (z_1+\alpha z_2) R(1-t^2)^2 \nonumber \\
&- \frac{1}{4} NR  m^2 (z_1+\alpha z_2) 
+ N R (z_1 p_1^2+\alpha z_2 p_2^2)+N D_x t^2,
\end{align}
where $z_1 = 4$, $z_2 = 2$, $\alpha=J^z/J$. For antiferromagnetic phase, the gap closes at $\Delta=\omega^{(1)}_{k_0} =0$ such that $\vec{k}_0 = (\pi,\pi,\pi)$.

The self-consistent equations for parameters $t, p_1, p_2$ and $\mu$ are obtained by minimizing the ground-state energy per site, given by:
\begin{align}
    \epsilon_0 = \frac{1}{N}\sum_{k} (\omega_k-\Lambda_k) +\frac{1}{N} C. 
\end{align}

To incorporate temperature dependence into the expectation values, we consider the Gibbs ensemble defined by 
\begin{align}\label{Gibbs0}
    G &= N \epsilon_0 - \frac{1}{\beta}\sum_k ln \left[1+n(\omega^{(1)}_k)\right] \nonumber \\
    &- \frac{1}{\beta}\sum_k ln \left[1+n(\omega^{(2)}_k)\right]
\end{align}
with $n(\omega^{(i)}_k) = 1/(e^{ \beta \omega^{(i)}_k }-1)$ and $\beta = \frac{1}{k_B T}$. 

Following the approach of Wang and Wang~\cite{WanglongPhysRevB,Wang}, for $D<D_C$,  we assume that a fraction of the bosonic excitations condense at \( k = \pi \). The self-consistent equations must then be solved by explicitly separating the Bose-Einstein condensate (BEC) fraction \( n_0(T) \), leading to the following set of equations(see appendix~\ref{condensation}):
\begin{equation}
t^2  = 2 -n_{h}(T)-\frac{1}{N}{\sum}^{\prime}_k \frac{\Lambda_k}{\omega_k}\left[1+n(\omega^{(1)}_k)+n(\omega^{(2)}_k)\right]
\label{eqt}
\end{equation}
\begin{align}
\mu  &= f_{\pi}  \left(1-\frac{\Delta_{\pi}}{\Lambda_{\pi}}   \right)n_h(T)  +  n_h(T) D_x  \nonumber \\ 
&+\frac{1}{N}{\sum}^{\prime}_k \frac{(\Lambda_k-\Delta_k)}{\omega_k}f_k\left[1+n(\omega^{(1)}_k)+n(\omega^{(2)}_k)\right] \nonumber\\
&+\frac{D_x}{N} {\sum}^{\prime}_k\frac{ \Lambda_k}{\omega_k} \left[1+n(\omega^{(1)}_k)+n(\omega^{(2)}_k)\right] 
\label{eqmu}
\end{align}
\begin{align}
p_1  &= \frac{1}{2} \frac{\Delta_{\pi}}{\Lambda_{\pi}}n_h(T)  \nonumber \\ 
&-\frac{1}{2N}{\sum}^{\prime}_k \frac{ \Delta_k \gamma^{(1)}_k}{\omega_k}\left[1+n(\omega^{(1)}_k)+n(\omega^{(2)}_k)\right]
\label{eqp1}
\end{align}
\begin{align}
p_2  &= \frac{1}{2} \frac{\Delta_{\pi}}{\Lambda_{\pi}}n_h(T)  \nonumber \\ 
&-\frac{1}{2N}{\sum}^{\prime}_k \frac{ \Delta_k \gamma^{(2)}_k}{\omega_k}\left[1+n(\omega^{(1)}_k)+n(\omega^{(2)}_k)\right]
\label{eqp2}
\end{align}
\begin{align}
m  &=   \frac{\omega_\pi}{\Lambda_{\pi}}n_h(T)+\frac{1}{N}{\sum}^{\prime}_k  \left[n(\omega^{(1)}_k)-n(\omega^{(2)}_k)\right]
\label{eqm}
\end{align}
here $f_k=z_1 \gamma^{(1)}_k+\alpha z_2 \gamma^{(2)}_k$.
Physically, only the gapless mode contributes to the staggered magnetization
$M_y = \sqrt{2t\, n_0(T)\left(1-\frac{\Lambda_{\pi}}{\Delta_{\pi}}\right)}$. 
The relation $\omega_{\pi}^{(1)} = 0$ 
defines the gapless magnon only within the ordered dome 
\(
h_{c1} < h < h_{c2}
\).
In this context, \(h_{c1}\) marks the first touching of the lower branch at 
\(k = \pi\) (onset of BEC), while \(h_{c2}\) marks the disappearance 
of the condensate and the onset of full polarization.


The staggered magnetization $M_y$ at $D < D_C$ is obtained from the transverse spin correlations. Although the local quantity $\langle (S_n^y)^2 \rangle$ remains finite even in the paramagnetic phase, true long-range order arises solely from the BEC-induced contribution to the correlation function $\langle S_i^y S_j^y \rangle$; accordingly, the finite staggered component originates from the magnon condensation.
\begin{figure}[t!]
\centering \includegraphics[width=1\linewidth]{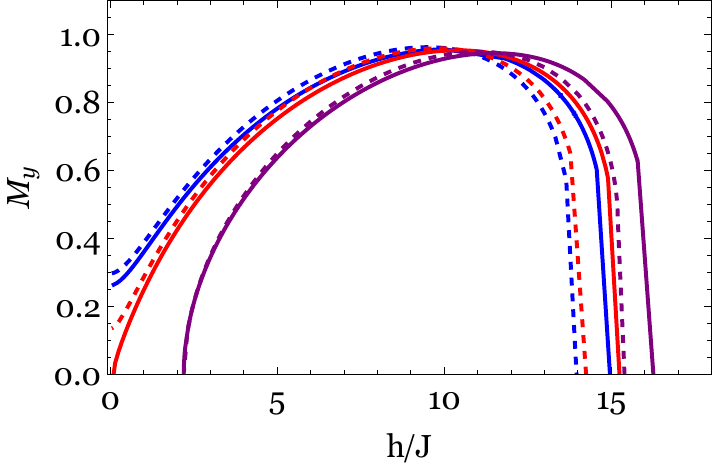}
\caption{Staggered magnetization $M_y$ as a function of $h/J$ at $J=1$, $T=0$, $R=1$ and 
$\alpha = 0.086$. Solid lines correspond to $D_x = 0$, while dashed lines represent $D_x = 0.5$. The curves for $D = 6.6$, $D = 6.9$, and $D = 8$ are shown in blue, red, and purple, respectively.}\label{fig:dome}
\end{figure}

\section{Results and discussion} \label{Results}

\paragraph{Bose-Einstein Condensation (BEC) and in-plane long-range order:} We solved \Cref{eqt,eqmu,eqp1,eqp2,eqm} numerically. The resulting values of 
$t$, $n_h(T)$, $\mu$, $p_1$, and $p_2$ determine both the staggered 
magnetization $M_y$ and the uniform magnetization $m$ as functions of 
$T$, $D$, $\alpha$, $D_x$, $h$, and $R$. For $D > D_c$, the ground state is a $D$-Large quantum paramagnet dominated by the 
$\ket{S^z = 0}$ component, so that $\langle (S_i^z)^2 \rangle \approx 0$, and the spin 
fluctuations are effectively confined to the $xy$ plane.

\paragraph{Staggered Magnetization Dome and Field-Induced BEC :} The Fig.~\ref{fig:dome} shows the staggered magnetization $M_y$ for $T=0$, $\alpha = 0.086$ and $R=1$ as
a function of $h/J$. Hereafter, we set $J=1$ (energy scale). The curves correspond to $D = 6.6$ (blue), $D = 6.9$ (red),
and $D = 8$ (purple). For each case, the solid lines denote $D_x = 0$, while the
dashed lines correspond to $D_x = 0.5$.

For $D = 6.6$, the staggered magnetization is finite at $h = 0$ for both 
$D_x = 0$ and $D_x = 0.5$. Upon increasing the field, $M_y$ develops a 
characteristic dome-shaped profile, and vanishes at 
$h_{c2}=14.94$ for $D_x = 0$ (blue solid line) and at 
$h_{c2}=13.94$ for $D_x = 0.5$ (blue dashed line). Beyond 
$h_{c2}$, the system becomes fully polarized along the external field.

In contrast, for $D = 6.9$, the staggered magnetization is zero at $h = 0$. As the field 
increases, $M_y$ again forms a dome as a function of $h$, 
closing at $h_{c2} = 15.24$ for $D_x = 0$ (solid red line) and at 
$h_{c2} = 14.24$ for $D_x = 0.5$ (dashed red line). Likewise, for $h \ge  h_{c2}$, the system is fully polarized.
\begin{figure}[t!]
\centering \includegraphics[width=1\linewidth]{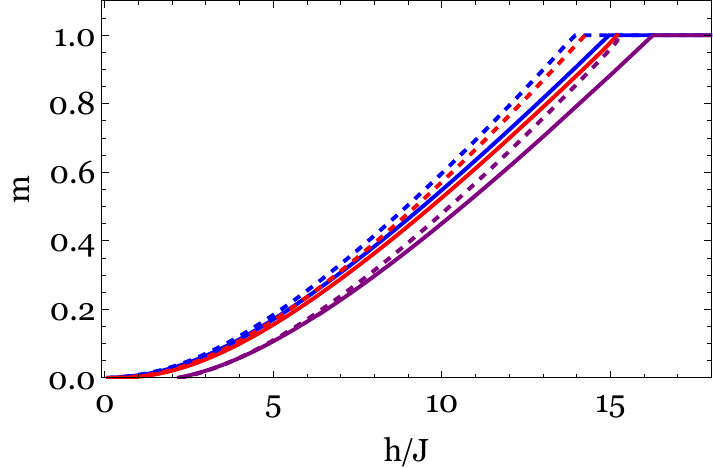}
\caption{Uniform magnetization $m$ as a function of $h/J$ at $J=1$, $T=0$, $R=1$, and 
$\alpha = 0.086$. Solid, dashed and colored lines follow the scheme of Fig.~\ref{fig:dome}.}\label{fig:magh}
\end{figure}

At $D = 8.0$, the staggered magnetization remains zero for 
$h \le h_{c1}=2.2$. As the field increases, $M_y$ exhibits the same dome-shaped behavior, closing at 
$h_{c2} = 16.34$ for $D_x = 0$ (purple solid line) and 
at $h_{c2} = 15.34$ for $D_x = 0.5$ (purple dashed line). 
Once more, for $h \ge h_{c2}$, the system is fully polarized.

\paragraph{The role of BEC:} 
In the presence of a magnetic field, the magnon spectrum splits into two branches (see~\Cref{ground,1stEx}). The energy of the lower branch, Eq.~(\ref{ground}), decreases with increasing field $h$ and vanishes at the critical value $h_{c1}$, where the excitations condense and the gap remains closed for $h \ge h_{c1}$. This condensation induces a uniform magnetization $m$ aligned with the field as well as a staggered component $M_y$ along the $y$-direction. At a second critical field $h_{c2}$, the uniform magnetization saturates, and the staggered component disappears. For $h > h_{c1}$, a critical temperature $T_C(h)$ exists below which the gap remains closed.

The field-induced dome reflects the Bose–Einstein condensation of the 
low-energy magnon mode $\omega^{(1)}_{\mathbf{k_0}}$ at 
$\mathbf{k_0} = (\pi,\pi,\pi)$. In this regime, $n_h(T)$ is finite, 
vanishing for $h \le h_{c1}$ and increasing steadily as the field is raised.
Inside the dome, the condensate density $n_h(T)$ is finite, and the corresponding order parameter is the transverse staggered magnetization $M_y$. 

Hence, the model in Eq.~(\ref{Model1}) exhibits a N\'eel-ordered phase, a quantum paramagnetic ($D$-Large) phase, and a field fully-polarized phase. 
The phase boundaries depend on the magnetic field as well as on the interplay between exchange couplings and anisotropies.
Figure~\ref{fig:magh} shows the uniform magnetization $m$ induced by the
external field $h$, obtained using the same parameters as in
Fig.~\ref{fig:dome}. 



For magnetic fields \( h \ge h_{c2} \), the ground state becomes fully polarized along the \( z \)-direction. 
At the critical field \( h_{c2} \), the variational parameter \( t \) vanishes, while \( n_{h_{c2}} = 1 \), and the chemical potential satisfies 
\(\mu = -(z_1 + \alpha z_2) + 2D_x\). Applying the gapless condition, we find that the critical point is given by (see Appendix \ref{AppH2c}.)
\begin{equation}\label{h2cmain}
h_{c2} = (z_1 + \alpha z_2)(1 + R) + D - 2D_x .
\end{equation}
\begin{figure}[t!]
\centering \includegraphics[width=0.95\linewidth]{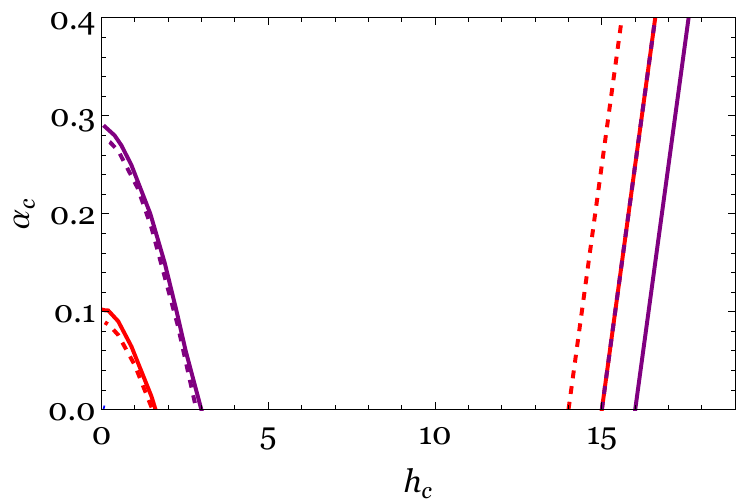}
\caption{Phase diagram of $\alpha_c$ versus the critical fields $h_{c1}$ and $h_{c2}$ for $R=1$. 
Red curves correspond to $D=7$, and purple curves to $D=8$. 
Solid lines denote $D_x=0$, while dashed lines indicate $D_x=0.5$. 
For $R<1$ (not shown), the phase boundary shifts to larger $\alpha_c$ at fixed $h$.}\label{alphaVshc}
\end{figure}
This compact expression clearly reveals how the anisotropies lower the critical field \( h_{c2} \), 
highlighting the influence of the in-plane anisotropy $D_x$ on the transition.
Equation~(\ref{h2cmain}) shows that the upper critical field is directly governed
by the microscopic anisotropies of the system. Since $h_{c2}$ depends
linearly on the exchange anisotropy $R$ and on the single--ion terms $D$ and
$D_x$, its experimental determination provides a sensitive probe of the
underlying symmetry-breaking mechanisms. In particular, deviations of $h_{c2}$
from the isotropic Heisenberg value $(z_1+\alpha z_2)$ make it possible to identify
whether the dominant anisotropy stems from exchange processes ($R\neq 1$),
from axial single--ion contributions ($D$), or from in--plane distortions
($D_x$). This renders Eq.~(\ref{h2cmain}) a practical diagnostic tool for extracting
magnetic anisotropies in layered and quasi--two--dimensional materials, where
high--field measurements are readily accessible.



\paragraph{Influence of interlayer interaction and anisotropies on the phase diagrams:}

In Fig.~\ref{alphaVshc} we show the phase diagram of $h$ versus $\alpha_c$ for $T=0$. 
The lines indicate the boundary separating the quantum paramagnetic 
($D$-Large) phase from the Néel-ordered phase with finite staggered 
magnetization $M_y$ in-plane. The region between $h_{c1}$ and $h_{c2}$ for a fixed value of $\alpha_c$ corresponds to an in-plane Néel-like phase, while the regimes $h < h_{c1}$ and $h > h_{c2}$ correspond to the quantum paramagnetic and fully polarized phases, respectively.
The red lines denote the phase boundary for 
$D = 6.9$, while the purple lines correspond to 
$D = 8$. On the other hand, the solid lines stand for $D_x=0$, and the dashed lines for $D_x=0.5$. Note that the effect of the anisotropy $D_x$ is most pronounced near $h_{c2}$; consistent with Eq.~(\ref{h2cmain}).

\begin{figure}[t!]
\centering \includegraphics[width=0.95\linewidth]{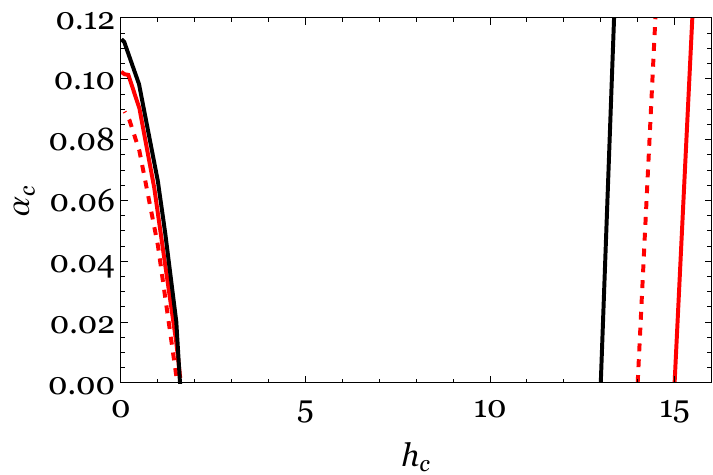}
\caption{ Phase diagram of $\alpha_c$ versus $h_{c1}$ for $D=7$. 
Red solid: $D_x=0$, $R=1$ (isotropic single-ion case). 
Red dashed: $D_x=0.5$, $R=1$ (single-ion distortion). 
Black solid: $D_x=0$, $R=0.5$. 
The curves illustrate how single-ion and exchange anisotropies shift the phase boundary; axes are labeled in the main panel.}\label{alphaVshcR}
\end{figure}

Fig.~\ref{alphaVshcR} shows the phase diagram of $\alpha_c$ as a function of the critical fields $h_{c1}$ and $h_{c2}$ at fixed $D=7$, highlighting the distinct effects of single-ion and exchange anisotropies. The solid red curve corresponds to the case $D_x=0$ and $R=1$.  
For $h<2$, it marks the phase boundary between the quantum paramagnetic regime 
and the N\'eel-like ordered state; in particular, for small $\alpha_c$ we obtain  $h_{c1}=1.6$.  
For $h>14$, the red solid curve identifies the transition from the N\'eel-like phase to the fully field-polarized regime, where $h_{c2}=15$.

Introducing a finite exchange anisotropy $R=0.5$ with $D_x=0$ (black solid line), we observe that for small $\alpha_c$ the lower critical field remains essentially unchanged at $h_{c1}\approx 1.6$, whereas the upper critical field is shifted to $h_{c2}\approx 13$ (black solid line).

Finally, for a finite single-ion distortion $D_x=0.5$ with $R=1$ (red dashed line), 
the lower critical field remains pinned at $h_{c1}\approx 1.6$ for small $\alpha_c$, 
while the upper critical field is reduced from $h_{c2}=15$ to $h_{c2}\approx 14$.

Together, these curves demonstrate how local (single-ion) and collective (exchange) anisotropies compete to reshape the stability region of the ordered phase in the $(\alpha_c,\, h_{c})$ plane.

Furthermore, it is worth emphasizing that the effects of anisotropies become considerably more pronounced in the upper critical field $h_{c2}$ than in the lower critical field $h_{c1}$ (see Fig.~\ref{alphaVshc} and Fig.~\ref{alphaVshcR}). This enhanced sensitivity near the fully polarized regime is particularly interesting, as it suggests that fields close to $h_{c2}$ can be used as a precise probe to characterize magnetic materials and to identify the nature of their anisotropies. Thus, $h_{c2}$ emerges as a key experimental quantity for resolving the nature of the anisotropies governing the magnetic behavior, although its interpretation must be complemented by additional experimental probes.

\paragraph{Experimental signatures:} The present results lead to several experimentally accessible signatures.  
First, the saturation field $h_{c2}$ provides the most direct probe of anisotropy. Its approximately linear dependence on both axial and in–plane single–ion terms, as well as on the exchange anisotropy, implies that high–field magnetization measurements can distinguish the dominant symmetry–breaking mechanism \cite{Zapf2014,Giamarchi2008}.  

Inelastic neutron scattering~\cite{INS} or THz spectroscopy \cite{Terahertz} offer complementary probes of 
the excitation spectrum. In the quantum–paramagnetic regime, these techniques 
can resolve the two magnon branches and track the softening of the lower mode 
upon approaching $h_{c1}$, as well as the rapid hardening of the upper branch 
when $h \to h_{c2}$.  

A further signature is encoded in the width of the BEC dome, $h_{c2}-h_{c1}$. While $h_{c1}$ is primarily determined by the exchange scale, the closure of the dome near $h_{c2}$ is highly sensitive to microscopic anisotropies, making the high–field regime an exceptionally sensitive diagnostic of subtle symmetry–breaking terms in layered spin–1 magnets.  
This provides a practical route for disentangling single–ion distortions from 
exchange anisotropy in real materials.

\section{Conclusion}\label{Conclusion}

We have shown that the high-field regime of layered spin-1 magnets provides 
a particularly sensitive window into the microscopic anisotropies that govern 
their spin dynamics. Within the SU(3) bond-operator framework, we obtained 
analytic expressions for the magnon spectrum and for the two critical fields 
$h_{c1}$ and $h_{c2}$ delimiting the field-induced ordered phase. 
Our results reveal a fundamental asymmetry between these two scales: 
while $h_{c1}$ is primarily controlled by the exchange couplings and is only weakly 
affected by single-ion terms, the saturation field $h_{c2}$ responds linearly 
and strongly to both exchange anisotropy and in-plane symmetry-breaking. 
This distinction makes $h_{c2}$ an experimentally accessible 
fingerprint of the dominant anisotropy mechanism. 

The framework developed here also clarifies how interlayer interactions, 
easy-plane anisotropy, and residual in--plane terms shape the size and 
location of the magnon Bose--Einstein condensation dome. 
These predictions directly motivate high-field probes, including magnetization, 
THz spectroscopy and inelastic neutron scattering, which can track the softening 
of the lower magnon branch at $h_{c1}$ and the rapid hardening of the upper branch 
as the field approaches $h_{c2}$.

Promising platforms for testing these results include several quasi--two--dimensional 
spin--1 Ni$^{2+}$ compounds in layered honeycomb or square geometries, 
such as BaNi$_2$V$_2$O$_8$, BaNi$_2$(PO$_4$)$_2$, BaNi$_2$(AsO$_4$)$_2$, 
La$_2$NiO$_4$, NiPS$_3$, and the prototypical BEC material DTN \cite{Knafo,Klyushina2017,PhysRevB.96.214428}.  
These systems exhibit well-characterized easy-plane anisotropy together with 
weaker in-plane symmetry breaking, making them ideal candidates for examining the 
anisotropy dependence of $h_{c2}$ and the evolution of the high-field magnon spectrum.  

Overall, our results establish a unified SU(3)–based framework with clear experimental signatures for identifying symmetry-breaking mechanisms in spin-1 magnets. They show that the competition among interlayer coupling, single-ion anisotropy, and exchange anisotropy governs the stability of the ordered phases in a highly nontrivial manner, while providing a direct route for confronting theory with high-field measurements in layered and quasi–two–dimensional materials.





\section*{Acknowledgments}

This work was partially supported by Fundação de Amparo a Pesquisa do Amazonas (FAPEAM) and Conselho Nacional de Desenvolvimento Cientifico e Tecnológico (CNPq).


\appendix

\section{SU(3) representation: explicit construction}\label{appSU3}

Starting with the eigenstates of $S^z$ operator $\ket{+1}$, $\ket{-1}$ and $\ket{0}$, we can write the following spin-1 states
\begin{align}
 \ket{x}
&= \frac{1}{\sqrt{2}}\big(\ket{+1}-\ket{-1}\big),\\
\ket{y}
&= \frac{i}{\sqrt{2}}\big(\ket{+1}+\ket{-1}\big),\\
\ket{z} &= \ket{0}.
\end{align}

Now, we define three bosonic operators $t_{\alpha}^\dagger$ with 
$\alpha = x,y,z$ such that $t_{x}^\dagger \ket{\text{vac}} = \ket{x}$,
$t_{y}^\dagger \ket{\text{vac}} = \ket{y}$ e
$t_{z}^\dagger \ket{\text{vac}} = \ket{z}$, with canonical commutation relations 
$\big[t_{\alpha},t^{\dagger}_{\beta}\big]=\delta_{\alpha\beta}$ and quantum vacuum state $\ket{\text{vac}}$. 

The spin-1 states $\ket{x}$, $\ket{y}$ and $\ket{z}$ can then be written as 
$\ket{\alpha}=t_\alpha^\dagger\ket{\text{vac}}$ ($\alpha=x,y,z$) and one readily verifies
\begin{equation}\label{projector}
t_{\beta}^{\dagger} t_{\delta} \ket{\gamma}
= \delta_{\delta\gamma}\, \ket{\beta}
\quad\Rightarrow\quad
t_{\beta}^{\dagger} t_{\delta} = \ket{\beta}\bra{\delta},
\end{equation}
i.e., $t_\beta^\dagger t_\delta$ acts as the matrix unit 
$\ket{\beta}\bra{\delta}$. Using Eq.~(\ref{projector}), the spin operators can be expressed as
\begin{align}\label{Sx}
S_x &= -i\big(\ket{y}\bra{z} - \ket{z}\bra{y}\big)
    = -i\big(t_y^\dagger t_z - t_z^\dagger t_y\big),
\end{align}
\begin{align}\label{Sy}
S_y &= -i\big(\ket{z}\bra{x} - \ket{x}\bra{z}\big)
    = -i\big(t_z^\dagger t_x - t_x^\dagger t_z\big),
\end{align}
\begin{align}\label{Sz}
S_z &= -i\big(\ket{x}\bra{y} - \ket{y}\bra{x}\big)
    = -i\big(t_x^\dagger t_y - t_y^\dagger t_x\big).
\end{align}
The \Cref{Sx,Sz}  can be written in the compact SU(3) form as 
\begin{equation}
S^\alpha = -i \sum_{\beta,\gamma}
\epsilon_{\alpha\beta\gamma}\, t_\beta^\dagger t_\gamma,
\end{equation}
where $\epsilon_{\alpha\beta\gamma}$ is the Levi--Civita tensor. These matrices reproduce the standard spin-1 representation of the angular momentum operators.

It is convenient to introduce the bosons $u$ and $d$ associated with the 
$S^z=\pm 1$ states via
\begin{equation}\label{relations}
d^{\dagger} = \frac{1}{\sqrt{2}}\big(t_x^{\dagger}-i t_y^{\dagger}\big),
\qquad
u^\dagger = -\frac{1}{\sqrt{2}}\big(t_x^\dagger + i t_y^\dagger\big),
\end{equation}
so that
\begin{equation}
u^{\dagger} \ket{\text{vac}} = \ket{+1},\quad
d^{\dagger} \ket{\text{vac}} = \ket{-1}
\end{equation}
Inverting the relations Eq.~(\ref{relations}), one finds
\begin{equation}\label{txty}
t_x = \frac{1}{\sqrt{2}}(d - u),\qquad
t_y = -\frac{i}{\sqrt{2}}(d + u),
\end{equation}
and similarly for the Hermitian conjugates. 
Substituting the Eq.~(\ref{txty}) into $S_x$, $S_y$ and $S^z$, and forming
$S^\pm = S_x \pm i S_y$ yields
\begin{align}
S^{+} &= S_x + i S_y \nonumber\\
      &= -i(t_y^\dagger t_z -t_z^\dagger t_y)
         + (t_z^\dagger t_x -t_x^\dagger t_z) \nonumber\\
      &= t_z^\dagger (t_x + i t_y) - (t_x^\dagger + i t_y^\dagger) t_z 
         \nonumber\\
      &= \sqrt{2}\big( t_z^{\dagger} d + u^{\dagger} t_z\big),\\[4pt]
S^{-} &= S_x - i S_y \nonumber\\
      &= -i(t_y^\dagger t_z -t_z^\dagger t_y)
         - (t_z^\dagger t_x -t_x^\dagger t_z) \nonumber\\
      &= t_z^\dagger (-t_x + i t_y)
         + (t_x^\dagger - i t_y^\dagger) t_z \nonumber\\
      &= \sqrt{2}\big(t_z^{\dagger} u + d^{\dagger} t_z\big).
\end{align}
and 
\begin{equation}
S^z = u^{\dagger}u - d^{\dagger}d.
\end{equation}
In all cases, the above expressions are understood to act within the 
physical subspace defined by the single-occupancy constraint
$u^\dagger u + d^\dagger d + t_z^\dagger t_z = 1$.

\section{Mean field Decoupling}\label{appendix:meanfield}

Decoupling intralayer:
\begin{align}
    (u_{\vec{r}}^\dagger & u_{\vec{r}} - d_{\vec{r}}^\dagger d_{\vec{r}})(u_{\vec{r}+\vec{\delta}}^\dagger u_{\vec{r}+\vec{\delta}} - d_{\vec{r}+\vec{\delta}}^\dagger d_{\vec{r}+\vec{\delta}}) \nonumber \\
    &=\frac{1}{2}(1-t^2+m)(u_{\vec{r}}^\dagger u_{\vec{r}}+ u_{\vec{r}+\vec{\delta}}^\dagger u_{\vec{r}+\vec{\delta}}) \nonumber \\ 
    &+\frac{1}{2}(1-t^2-m)(d_{\vec{r}}^\dagger d_{\vec{r}}+ d_{\vec{r}+\vec{\delta}}^\dagger d_{\vec{r}+\vec{\delta}}) \nonumber \\
    &-p_1(u_{\vec{r}} d_{\vec{r}+\vec{\delta}} + d_{\vec{r}} u_{\vec{r}+\vec{\delta}} +H. c. )-\frac{1}{2}(1-t^2)^2 \nonumber \\
    &-\frac{1}{2} m^2 +2p_1^2
\end{align}
with $p_1=\langle u_{\vec{r}}d_{\vec{r}+\vec{\delta}} \rangle=\langle u^{\dagger}_{\vec{r}}d^{\dagger}_{\vec{r}+\vec{\delta}} \rangle$, $m=\langle u^{\dagger}_{\vec{r}}u_{\vec{r}} \rangle-\langle d^{\dagger}_{\vec{r}}d_{\vec{r}} \rangle$

Decoupling interlayer:
\begin{align}
    (u_{\vec{r}}^\dagger & u_{\vec{r}} - d_r^\dagger d_{\vec{r}})(u_{\vec{r}+\vec{\xi}}^\dagger u_{\vec{r}+\vec{\xi}} - d_{\vec{r}+\vec{\xi}}^\dagger d_{\vec{r}+\vec{\xi}}) \nonumber \\
    &=\frac{1}{2}(1-t^2+m)(u_{\vec{r}}^\dagger u_{\vec{r}}+ u_{\vec{r}+\vec{\xi}}^\dagger u_{\vec{r}+\vec{\xi}}) \nonumber \\ 
    &+\frac{1}{2}(1-t^2-m)(d_{\vec{r}}^\dagger d_{\vec{r}}+ d_{\vec{r}+\vec{\xi}}^\dagger d_{\vec{r}+\vec{\xi}}) \nonumber \\
    &-p_1(u_rd_{\vec{r}+\vec{\xi}} + d_ru_{\vec{r}+\vec{\xi}} +H. c. )-\frac{1}{2}(1-t^2)^2 \nonumber \\
    &-\frac{1}{2} m^2 +2p_2^2
\end{align}
with $p_1=\langle u_{\vec{r}}d_{\vec{r}+\vec{\xi}} \rangle=\langle u^{\dagger}_{\vec{r}}d^{\dagger}_{\vec{r}+\vec{\xi}} \rangle$, $p_2=\langle u_{\vec{r}}d_{\vec{r}+\vec{\xi}} \rangle=\langle u^{\dagger}_{\vec{r}}d^{\dagger}_{\vec{r}+\vec{\xi}} \rangle$, $m=\langle u^{\dagger}_{\vec{r}}u_{\vec{r}} \rangle-\langle d^{\dagger}_{\vec{r}}d_{\vec{r}} \rangle$. The present mean-field decoupling retains the correlations between the operators $u$ and $d$.

\section{Self-consistent equations}\label{condensation}

For $D>D_c$, the minimization of the Gibb`s energy Eq.~(\ref{Gibbs0}) leads to the self-consistent equations 

\begin{equation}
t^2  = 2 -\frac{1}{N}\sum_k \frac{\Lambda_k}{\omega_k}[1+n(\omega^{(1)}_k)+n(\omega^{(2)}_k)]
\end{equation}

\begin{align}
\mu  &= \frac{1}{N}\sum_k \frac{(\Lambda_k-\Delta_k)}{\omega_k}f_k[1+n(\omega^{(1)}_k)+n(\omega^{(2)}_k)]
\end{align}

\begin{align}
p_1  &= -\frac{1}{2N}\sum_k \frac{ \Delta_k \gamma^{(1)}_k}{\omega_k}[1+n(\omega^{(1)}_k)+n(\omega^{(2)}_k)]
\end{align}

\begin{align}
p_2  &= -\frac{1}{2N}\sum_k \frac{ \Delta_k \gamma^{(2)}_k}{\omega_k}[1+n(\omega^{(1)}_k)+n(\omega^{(2)}_k)]
\end{align}

\begin{align}
m  &=  \frac{1}{N}\sum_k  [n(\omega^{(1)}_k)-n(\omega^{(2)}_k)]
\end{align}
where $f_k=z_1 \gamma^{(1)}_k+\alpha z_2 \gamma^{(2)}_k$ with $\gamma^{(1)}_k=\frac{1}{2}(\cos(k_x)+\cos(k_y))$,  $\gamma^{(2)}_k=\cos(k_z)$, $z_1=4$ and $z_2=2$.

Below $D_c$, the lower branch $\omega_{\mathbf{k}}^{(1)}$ becomes gapless 
at $\vec{k}_0=(\pi,\pi,\pi)$ and its occupation acquires a macroscopic component. 
To treat this mode, we separate explicitly the $\vec{k}_0$ contribution 
from the momentum sums. In the constraint equation for $t^2$, the singular 
term is proportional to $(\Lambda_\pi/\omega_\pi)\,n(\omega_\pi^{(1)})/N$, 
which we replace by a condensate density $n_h(T)$. The remaining factor 
$(\Lambda_\pi/\omega_\pi)[1+n(\omega_\pi^{(2)})]/N$ stays finite and is 
absorbed into the regular part of the momentum sums. Hence,
\begin{equation}
t^2  = 2 -n_{h}(T)-\frac{1}{N}{\sum}^{\prime}_k \frac{\Lambda_k}{\omega_k}[1+n(\omega^{(1)}_k)+n(\omega^{(2)}_k)]
\end{equation}\label{eqt2}
\begin{align}
\mu  &= f_{\pi}  \left(1-\frac{\Delta_{\pi}}{\Lambda_{\pi}}   \right)n_h(T)  +  n_h(T) D_x  \nonumber \\ 
&+\frac{1}{N}{\sum}^{\prime}_k \frac{(\Lambda_k-\Delta_k)}{\omega_k}f_k[1+n(\omega^{(1)}_k)+n(\omega^{(2)}_k)] \nonumber\\
&+\frac{D_x}{N} {\sum}^{\prime}_k\frac{ \Lambda_k}{\omega_k} [1+n(\omega^{(1)}_k)+n(\omega^{(2)}_k)] 
\end{align}\label{eqmu2}
\begin{align}
p_1  &= \frac{1}{2} \frac{\Delta_{\pi}}{\Lambda_{\pi}}n_h(T)  \nonumber \\ 
&-\frac{1}{2N}{\sum}^{\prime}_k \frac{ \Delta_k \gamma^{(1)}_k}{\omega_k}[1+n(\omega^{(1)}_k)+n(\omega^{(2)}_k)]
\end{align}\label{eqp1b}
\begin{align}
p_2  &= \frac{1}{2} \frac{\Delta_{\pi}}{\Lambda_{\pi}}n_h(T)  \nonumber \\ 
&-\frac{1}{2N}{\sum}^{\prime}_k \frac{ \Delta_k \gamma^{(2)}_k}{\omega_k}[1+n(\omega^{(1)}_k)+n(\omega^{(2)}_k)]
\end{align}\label{eqp2b}
\begin{align}
m  &=   \frac{\omega_\pi}{\Lambda_{\pi}}n_h(T)+\frac{1}{N}{\sum}^{\prime}_k  [n(\omega^{(1)}_k)-n(\omega^{(2)}_k)]
\end{align}\label{eqm2}
where $f_k=z_1 \gamma^{(1)}_k+\alpha z_2 \gamma^{(2)}_k$.

\section{Polarized phase}\label{AppH2c}

At $h = h_{c2}$ the system becomes fully polarized along the external field. 
For $h > h_{c2}$ the variational parameters take the saturated values 
$t = 0$, $p_1 = p_2 = 0$, and $m = 1$. 
From Eq.~(\ref{eqt}) we then obtain $n_h(T)=1$, and from Eq.~(\ref{eqmu}) the chemical potential 
reduces to 
\[
\mu = -(z_1+\alpha z_2) + 2D_x .
\]
In this limit the excitation energy becomes
\[
\omega_{\pi} = -\mu + D + \tfrac{1}{2} Z R .
\]

The gapless condition is therefore
\begin{equation}
\label{gaplessconditions}
\omega_{\pi} - h_{c2} + \tfrac{1}{2} R (z_1 + \alpha z_2) = 0,
\end{equation}
and substituting the saturated values into Eq.~(\ref{gaplessconditions}) yields
\begin{equation}\label{h2c}
h_{c2} = (z_1 + \alpha z_2)(1 + R) + D - 2D_x .
\end{equation}
\vspace{0.5cm}

\section{ Effective terms at low energy}\label{Approximations}

In the field-induced ordered phase, only the $u$–magnon branch becomes soft and condenses, while the opposite-flavour $d$ branch remains gapped by an energy scale $\Delta ^\prime \sim\mathcal{O}(h)$. The $D_x$ term generates both density-like contributions, $u_{\mathbf{k}}^{\dagger}u_{\mathbf{k}}+ d_{\mathbf{k}}^{\dagger}d_{\mathbf{k}}$, and anomalous terms of the form $u_{\mathbf{k}} d_{-\mathbf{k}} + u_{\mathbf{k}}^{\dagger} d_{-\mathbf{k}}^{\dagger}$. The anomalous part couples the critical $u$ magnons to virtual two-magnon processes involving the gapped $d$ branch. Treating this coupling perturbatively yields a second-order self-energy correction,
\[
\Sigma_{u}(\mathbf{k},\omega) 
    \sim 
    \frac{D_x^{\,2}}{\omega - \Delta ^\prime}
    \;\;\Rightarrow\;\;
    \delta\varepsilon_{u}(\mathbf{k}) 
    \sim \mathcal{O}\!\left(\frac{D_x^{2}}{ \Delta ^\prime}\right),
\]
which is parametrically small as long as $D_x\ll \Delta ^\prime$ ( e.g. in this paper, we considered  $\frac{D_x}{\Delta^{\prime}} \sim 10^{-2}$). Thus the anomalous terms contribute only a subleading renormalization to the $u$-magnon dispersion, while the dominant effect of $D_x$ is already captured by the density-like terms. For this reason, in our analysis of the field-polarized ordered phase we neglect these anomalous contributions and retain only the diagonal part of $H_{D_x}$.

\bibliographystyle{IEEEtran}
\bibliography{refs}

\end{document}